\makeatletter
\input{filecontents.sty}
\makeatother

\documentclass[%
titlepage,
titlepageabstract,
secnumarabic,
showpacs,
twocolumn,
twoside,
floatfix,
a4paper,
superscriptaddress,
groupedaddress,
amsmath,amssymb,amsfonts,amsbsy,aps]{preprint}

\makeatletter
\def\lrbox#1{%
  \edef\reserved@a{%
    \endgroup
    \global\setbox#1\hbox{%
      \begingroup\aftergroup}%
    \def\noexpand\@currenvir{\@currenvir}%
    \def\noexpand\@currenvline{\on@line}}%
  \reserved@a
  \@endpefalse
  \color@setgroup
  \ignorespaces}
\def\endlrbox{\unskip\color@endgroup}
\makeatother

\renewcommand\documentclass[2][]{}
\newcommand{\changefont}[3]{}
\newcommand{\journal}[1]{}
\newbox\mybox
\newenvironment{frontmatter}{}{\maketitle}
\def\sep{\unskip, }%
\let\ead=\email
\let\PACS=\pacs
\newenvironment{keyword}{\begin{lrbox}{\mybox}\footnotesize}{\end{lrbox}\keywords{\usebox{\mybox}}}
\AtBeginDocument{%
  \IfFileExists{date.tex}{%
    \input{date.tex}

  }
}
\renewcommand\documentclass[2][]{}
\documentclass[final,5p,twocolumn]{elsarticle}

\usepackage{graphicx}
\graphicspath{{figures/}}
\usepackage{amsmath}
\usepackage{amssymb}
\usepackage{amsthm}
\usepackage{microtype}
\usepackage{txfonts}
\usepackage{mdwlist}
\usepackage[
breaklinks=true,dvipdfmx,
pdfauthor={Heiko Bauke, and Christoph H. Keitel},
pdftitle={Accelerating the Fourier split operator method via graphics processing units}
]{hyperref}
\renewcommand{\vec}[1]{\boldsymbol{#1}}
\newcommand{\D}{\mathrm{d}}
\newcommand{\I}{\mathrm{i}}
\newcommand{\e}{\mathrm{e}}
\newcommand*{\op}[1]{{\hat{#1}}}
\newcommand*{\ft}[1]{\mathcal{F}\left\{#1\right\}}
\newcommand*{\ftinv}[1]{\mathcal{F}^{-1}\left\{#1\right\}}

\journal{Computer Physics Communications}

\begin{document}

\begin{frontmatter}
  \title{Accelerating the Fourier split operator method via graphics processing units}
  \author{Heiko Bauke}
  \ead{bauke@mpi-hd.mpg.de}
  \author{Christoph H. Keitel}
  \ead{keitel@mpi-hd.mpg.de}
  \address{Max-Planck-Institut f{\"u}r Kernphysik, Saupfercheckweg~1,
    69117~Heidelberg, Germany}
  \begin{abstract}
    Current generations of graphics processing units have turned into
    highly parallel devices with general computing capabilities. Thus,
    graphics processing units may be utilized, for example, to solve
    time dependent partial differential equations by the Fourier split
    operator method. In this contribution, we demonstrate that graphics
    processing units are capable to calculate fast Fourier transforms
    much more efficiently than traditional central processing
    units. Thus, graphics processing units render efficient
    implementations of the Fourier split operator method
    possible. Performance gains of more than an order of magnitude as
    compared to implementations for traditional central processing units
    are reached in the solution of the time dependent Schr{\"o}dinger
    equation and the time dependent Dirac equation.
  \end{abstract}
\end{frontmatter}

\section{Introduction}
\label{sec:intro}

For several decades, users and developers of high performance computing
applications could trust on Moore's law. The number of transistors that
can be placed inexpensively on an integrated circuit has doubled
approximately every two years. This exponential growth provided the
basis for an ever-increasing computing power of modern central
processing units (CPUs). Thus, high performance computing applications
became faster by just trusting on Moore's law and waiting for faster
CPUs.  Thanks to Moore's law, today's desktop computers supply the
computational power of the last decade's super-computers.

Transistor counts still continue to grow exponentially. Hardware
manufacturers, however, no longer make single computing units more
complex and faster. Clock rates and execution optimizations have reached
their limits. Now hardware manufacturers use the plethora of transistors
to put many computing units on a single chip manufacturing highly
parallel computing architectures. There are two major directions of
development, multicore architectures with a few (typically two to a few
dozen in the near future) general purpose computing units and dedicated
accelerators with several hundred or more computing units with reduced
capabilities each as compared to traditional CPUs
\cite{Kirk:Hwu:2010:Programming_MPP,Kindratenko:etal:2010:_HPC_Accelerators}.
These accelerators may be implemented in field-programmable gate arrays
or may be custom built systems as the Cell Broadband Engine Architecture
or the ClearSpeed\texttrademark\ CSX700 Processor, for example. Graphics
processing units (GPUs) with general purpose computing capabilities
brought accelerated computing to the mass market.  Employing GPUs to
perform computations that are traditionally handled by CPUs is commonly
referred to as GPU computing \cite{Owens:etal:2008:GPU_Computing}.

Parallel architectures are not new in the high performance computing
community. In the past, many high performance computing applications
employed various kinds of parallel computing architectures which had
been the high-end computing systems of their time. Today, however, even
low-end consumer computers are parallel systems. The new emerging
ubiquitous parallel architectures indicate a silent paradigm change in
computing or as Herb Sutter \cite{Sutter:2005:free_lunch} pointed out
``The free lunch is over.'' In order to exploit the power of modern
hardware architectures it is required to write parallel applications.
The term ``computing'' becomes synonymous to ``parallel computing.''

The purpose of this contribution is to evaluate the Fermi GPU computing
architecture and to demonstrate how an implementation of the Fourier
split operator method on GPUs may boost the performance of the numerical
propagation of time dependent partial differential equations. The
remainder of this paper is organized as follows. In
section~\ref{sec:Fourier_split} we give a generic description of the
Fourier split operator method and show how to apply it to the time
dependent Schr{\"o}dinger equation and the time dependent Dirac
equation. Section~\ref{sec:implementation} gives a very short
introduction to the CUDA architecture and describes our GPU
implementation of the Fourier split operator method while
section~\ref{sec:performance} presents performance comparisons. In
section~\ref{sec:applications} we illustrate some applications of the
Fourier split operator method.

\section{The Fourier split operator method}
\label{sec:Fourier_split}

Let us consider the Cauchy type initial value problem problem
\begin{subequations}
  \label{eq:Cauchy}
  \begin{equation}
    \frac{\partial w(\vec x, t)}{\partial t} =
    \op A(t) w(\vec x, t)
  \end{equation}
  with some possibly time dependent differential operator $\op A(t)$
  acting on the coordinates $\vec x =(x_1,x_2,\dots,x_n)$ and with the
  initial condition
  \begin{equation}
    w(\vec x, 0) = w_0(\vec x) \,.
  \end{equation}
\end{subequations}
The function $w(\vec x, t)$ depends on $\vec x$ and the time coordinate
$t$ and may be real or complex and possibly vector valued. The Cauchy
problem (\ref{eq:Cauchy}) includes, for example, the diffusion equation,
the Focker-Planck equation \cite{Risken:1989:Focker-Planck}, the
current-free Maxwell equations \cite{Harshawardhan:etal:2000:Maxwell},
the Schr{\"o}dinger equation, the Pauli equation, the Dirac equation
\cite{Thaller:1992:Dirac}, as well as the Klein-Gordon equation in the
Feshbach-Villars representation
\cite{Feshbach:Villars:1954:Relativistic_Wave_Mechanics} or the
Black-Scholes options pricing equation.

Using Dyson's time ordering operator $\hat T$, the formal solution of
(\ref{eq:Cauchy}) is given by
\begin{equation}
  w(\vec x, t)=\op U(t, 0) w(\vec x, 0)
\end{equation}
with the time-evolution operator
\cite{Pechukas:Light:1966:Time-Displacement,Fattorini:1985:Cauchy,Risken:1989:Focker-Planck}
\begin{equation}
  \label{eq:U_op}
  \op U(t_2, t_1) =
  \op T\exp\left(\int_{t_1}^{t_2} \op A(t') \,\D t' \right)
\end{equation}
that effects the function's evolution from time $t_1$ to time~$t_2$. For
some highly symmetric operators $\op A(t)$ in (\ref{eq:Cauchy}),
analytic expressions of the time-evolution operator (\ref{eq:U_op}) can
be calculated. However, investigations of many relevant problems require
numerical methods, for example, the Fourier split operator method, which
we will shortly describe in the following paragraphs.

\subsection{General outline of the Fourier split operator method}
\label{sec:Fourier_split_gen}

Fleck et~al.\ \cite{Fleck:etal:1976:laser_propagation} introduced the
Fourier split operator method as an explicit time stepping scheme for
the solution of the time dependent scalar Maxwell wave equation in
Fresnel approximation. The method is applicable to Cauchy problems
(\ref{eq:Cauchy}) with a differential operator $\op A(t)$ that has the
property that it may be separated into a sum of two operators
\begin{equation}
  \label{eq:sum_A}
  \op A(t) = \op{A}_1(t) + \op{A}_2(t) 
\end{equation}
such that $\op{A}_1(t)$ can be easily diagonalized in real space while
$\op{A}_2(t)$ can be easily diagonalized in Fourier space; a requirement
that is fulfilled by many partial differential equations of relevance.
Thus, shortly after Feit et~al.\ \cite{Feit:etal:1982:spectral_method}
solved the Schr{\"o}dinger equation with a time independent Hamiltonian
numerically by the Fourier split operator method, it became a standard
tool for the propagation of quantum mechanical wave equations.  Later,
the method had been generalized to the Schr{\"o}dinger equation with a
time dependent Hamiltonian
\cite{Bandrauk:Shen:1997:coupled_time-dependent_SE} and it was applied
to other equations, for example, the Dirac equation
\cite{Braun:etal:1999:Numerical_approach,Mocken:Keitel:2004:Quantum_dynamics,Mocken:Keitel:2008:FFT-split-operator,Hu:Keitel:2001:Dynamics},
the time dependent Gross-Pitaevskii equation
\cite{Javanainen:Ruostekoski:2006:Symbolic_calculation}, the non-linear
Schr{\"o}dinger equation
\cite{Muslu:Erbay:2005:Higher-order_split-step}, and the time dependent
Maxwell equations for electromagnetic waves in random dielectric media
\cite{Harshawardhan:etal:2000:Maxwell}.

The central idea of the Fourier split operator method is to approximate
the operator (\ref{eq:U_op}) by a product of operators that are diagonal
either in real space or in momentum space.  Let $\op O(t)$ denote some
possibly time dependent operator and define the operator
\begin{equation}
  \label{eq:U_O_op}
  \op U_{\op O}(t_2, t_1, \delta) =
  \exp\left( \delta \int_{t_1}^{t_2} \op O(t') \,\D t' \right)\,,
\end{equation}
that depends on the times $t_1$ and $t_2$ and the auxiliary
parameter~$\delta$ and let $\op {\tilde{ U}}_{\op O}(t_2, t_1, \delta)$
denote the operator (\ref{eq:U_O_op}) in Fourier space. Expanding $\op
U\left(t + \tau, t\right)$ to the third order in $\tau$ and provided
that the operator $\op A(t)$ has the form (\ref{eq:sum_A}), the
time-evolution operator (\ref{eq:U_op}) can be factorized
\cite{Pechukas:Light:1966:Time-Displacement} into
\begin{multline}
  \label{eq:U_split}
  \op U\left(t + \tau, t\right) = 
  \exp\left(\int_{t}^{t+\tau} \op A(t') \,\D t' \right) + O\left(\tau^3\right) = \\ 
  \op U_{\op{A}_1}\left(t + \tau, t, \tfrac12\right)
  \op U_{\op{A}_2}\left(t + \tau, t, 1\right)
  \op U_{\op{A}_1}\left(t + \tau, t, \tfrac12\right) + O\left(\tau^3\right)\,.
\end{multline}
Neglecting terms of order $O\left(\tau^3\right)$, equation
(\ref{eq:U_split}) gives an explicit second order accurate time-stepping
scheme for the propagation of the function $w(\vec x, t)$
\begin{multline}
  \label{eq:step_t_to_t+dt}
  w(\vec x, t+\tau) = \\
  \op U_{\op{A}_1}\left(t + \tau, t, \tfrac12\right)
  \op U_{\op{A}_2}\left(t + \tau, t, 1\right)
  \op U_{\op{A}_1}\left(t + \tau, t, \tfrac12\right)w(\vec x, t)
  + O\left(\tau^3\right)
  \,.
\end{multline}
This scheme translates the difficulty of calculating the action of
operator (\ref{eq:U_op}) to the task of calculating the action of
(\ref{eq:U_O_op}) for $\op O\equiv \op{A}_1$ and $\op O\equiv \op{A}_2$,
respectively. Strang \cite{Strang:1968:Difference_Schemes} utilized the
splitting (\ref{eq:step_t_to_t+dt}) to calculate the action of
(\ref{eq:U_O_op}) in real space by a finite differences scheme. For many
Cauchy problems (\ref{eq:Cauchy}), however, one can find a splitting
(\ref{eq:sum_A}) such that the operator $\op U_{\op{A}_1}\left(t+\tau, t,
  \delta\right)$ is diagonal in real space and $\op U_{\op
  A_2}\left(t+\tau, t, \delta\right)$ is diagonal in Fourier
space. Thus, the calculation of these operators becomes feasible in the
appropriate space and (\ref{eq:step_t_to_t+dt}) is calculated via
\begin{multline}
  \label{eq:step_t_to_t+dt_F}
  w(\vec x, t+\tau) = \\
  \op U_{\op{A}_1}\left(t + \tau, t, \tfrac12\right)
  \ftinv{\vphantom{\bigg]}
    \op {\tilde U}_{\op{A}_2}\left(t + \tau, t, 1\right)
    \ft{\op U_{\op{A}_1}\left(t + \tau, t, \tfrac12\right)w(\vec x, t)}
  } \\
  + O\left(\tau^3\right) \,.
\end{multline}
The expression $\ft{\cdot}$ in (\ref{eq:step_t_to_t+dt_F}) denotes the
Fourier transform of the argument and $\ftinv{\cdot}$ the inverse
Fourier transform.

In a computer implementation of the Fourier split operator method, the
function $w(\vec x, t)$ is discretized on a rectangular regular lattice
of $N$ points and the continuous Fourier transform is approximated by a
discrete Fourier transform.  The computational complexity of propagating
the function $w(\vec x, t)$ from time $t$ to time $t+\tau$ is dominated
by the transformation into Fourier space and back into real space. If
these transforms are accomplished by the fast Fourier transform the
computation of an elementary step of the Fourier split operator method
takes $O(N\log N)$ operations.

\subsection{Fourier split operator method for the Schr{\"o}dinger equation}
\label{sec:Fourier_split_SE}

The Schr{\"o}dinger equation is an equation of motion for a complex-valued
scalar wave function $\Psi(\vec x, t)$ evolving in a
\mbox{$d$-dimensional} space $\vec x$ and in time $t$. In its most general
form, it describes a non-relativistic spin zero particle of mass $m$ and
charge $q$ in the electromagnetic potentials $\phi(\vec x, t)$ and $\vec
A(\vec x, t)$. The Schr{\"o}dinger equation is invariant under Galilean
transformation and reads
\begin{multline}
  \label{eq:SE}
  \I\hbar\frac{\partial \Psi(\vec x, t)}{\partial t} = 
  \\
  \left(
    \frac{1}{2m}\left( -\hbar\I\nabla - q \vec A(\vec x, t) \right)^2
    + q \phi(\vec x, t)
  \right)\Psi(\vec x, t) \,.
\end{multline}

In order to apply the Fourier split operator method for propagating
$\Psi(\vec x, t)$ under the effect of (\ref{eq:SE}), we have to restrict
the vector potential to homogeneous fields $\vec A(\vec x, t)=\vec
A(t)$; an approximation that is known as the dipole approximation.
Splitting the Hamiltonian (\ref{eq:SE}) in dipole approximation into a
potential energy term and a kinetic energy term
\begin{subequations}
  \begin{align}
    \label{eq:SE_V}
    {\I\hbar}\op{A}_1 
    & = q \phi(\vec x, t)\,, \\
    \label{eq:SE_K}
    {\I\hbar}\op{A}_2 
    & =
    \frac{1}{2m}\left( -\hbar\I\nabla - q \vec A(t) \right)^2
  \end{align}
\end{subequations}
separates the spatial dependent parts from spatial derivatives, which
makes the operator $\op U_{\op{A}_1}\left(t+\tau, t, \delta\right)$
diagonal in real space and $\op U_{\op{A}_2}\left(t+\tau, t,
  \delta\right)$ diagonal in momentum space, respectively. The action of
$\op U_{\op{A}_1}\left(t+\tau, t, \delta\right)$ on a real space wave
function is given by 
\begin{equation}
  \op U_{\op{A}_1}\left(t+\tau, t, \delta\right)\Psi(\vec x, t) =
  \exp\left(
    -\delta\frac{\I}{\hbar}\int_t^{t+\tau} q \phi(\vec x, t') \,\D t'
  \right)\Psi(\vec x, t)
\end{equation}
and the action of the Fourier space operator $\op{\tilde U}_{\op
  A_2}\left(t+\tau, t, \delta\right)$ to a wave function in Fourier
space 
\begin{multline}
  \tilde\Psi(\vec p, t) = 
  \ft{\Psi(\vec x, t)} = \\
  \frac{1}{(2\pi \hbar^2)^{d/2}}\int \Psi(\vec x, t)\exp\left(
    -\I \vec p\cdot\vec x/\hbar
  \right)\,\D^d x
\end{multline}
reads
\begin{multline}
  \op{\tilde U}_{\op{A}_2}\left(t+\tau, t, \delta\right)
  \tilde\Psi(\vec p, t) = \\
  \exp\left(
    -\delta\frac{\I}{\hbar}\int_t^{t+\tau}  \frac{1}{2m}\left( 
      \vec p - q\vec A(t') 
    \right)^2\,\D t'
  \right)\tilde\Psi(\vec p, t)\,.
\end{multline}

Note that it is crucial for the application of the Fourier split
operator method that the vector potential $\vec A(t)$ does not depend on
the spatial coordinate~$\vec x$. The expansion of the Hamiltonian of the
Schr{\"o}dinger equation (\ref{eq:SE}) for a particle in an arbitrary vector
potential $\vec A(\vec x, t)$ contains the term $({\I q\hbar}/{m})\vec
A(\vec x, t)\cdot\nabla$, that is spatially dependent and contains
spatial derivatives, too, coupling momentum and coordinate
space. Coupling between momentum and coordinate space is absent for
vector potentials in dipole approximation but also, for example, for the
vector potential of a linearly polarized plane wave with $\vec A(\vec x,
t)=(A_1(x_3, t), 0, 0)$ \cite{Hu:Keitel:2001:Dynamics}.

\subsection{Fourier split operator method for the Dirac equation}
\label{sec:Fourier_split_DE}

In contrast to the Schr{\"o}dinger equation, the Dirac equation
\cite{Thaller:1992:Dirac} describes a relativistic spin half particle
and it is invariant under Lorentz transformation. A Dirac wave function
$\Psi(\vec x, t)$ is a two component (for one-dimen\-sional systems) or
four component (for two- or three-dimen\-sional systems) complex-valued
vector function.  The Dirac equation for a particle of mass $m$ and
charge $q$ moving in the electromagnetic potentials $\phi(\vec x, t)$
and $\vec A(\vec x, t)$ is given by
\begin{multline}
  \label{eq:DE}
  \I\hbar\frac{\partial \Psi(\vec x, t)}{\partial t} =
  \\ 
  \left(
    c\sum_{i=1}^d
    \alpha_i\left(
      -\I\hbar\frac{\partial}{\partial r_i} - q A_i(\vec x, t)
    \right) + q \phi(\vec x, t) + mc^2\beta
  \right)\Psi(\vec x, t)
\end{multline}
with the matrices $\alpha_i$ and $\beta$ and the speed of light $c$ in
vacuum and $A_i(\vec x, t)$ denoting the $i$th component of the vector
potential $\vec A(\vec x, t)$.  The matrices $\alpha_i$, $\beta$ obey
the algebra
\begin{equation}
  \label{eq:Dirac_algebra}
  \alpha_i^2 = \beta^2 = 1\,,\quad
  \alpha_i\alpha_k + \alpha_k\alpha_i = 2\delta_{i,k}\,,\quad
  \alpha_i\beta + \beta\alpha_i = 0\,.
\end{equation}
This algebra determines the matrices $\alpha_i$ and $\beta$ only up to
unitary transforms. In numerical applications, we adopted the so-called
Dirac representation with
\begin{equation}
  \label{eq:Dirac_repr}
  \begin{aligned}
    \alpha_1 & =
    \begin{pmatrix}
      0 & 0 & 0 & 1 \\
      0 & 0 & 1 & 0 \\
      0 & 1 & 0 & 0 \\
      1 & 0 & 0 & 0 
    \end{pmatrix}\,,
    &
    \alpha_2 & =
    \begin{pmatrix}
      0 & 0 & 0 & -\I \\
      0 & 0 & \I & 0 \\
      0 & -\I & 0 & 0 \\
      \I & 0 & 0 & 0 
    \end{pmatrix}\,,
    \\
    \alpha_3 & =
    \begin{pmatrix}
      0 & 0 & 1 & 0 \\
      0 & 0 & 0 & -1 \\
      1 & 0 & 0 & 0 \\
      0 & -1 & 0 & 0 
    \end{pmatrix}\,,
    &
    \beta & =
    \begin{pmatrix}
      1 & 0 & 0 & 0 \\
      0 & 1 & 0 & 0 \\
      0 & 0 & -1 & 0 \\
      0 & 0 & 0 & -1
    \end{pmatrix}\,.
  \end{aligned}
\end{equation}
For one-dimensional systems, the Dirac representation reduces to
\begin{equation}
  \begin{aligned}
    \alpha_1 & =
    \begin{pmatrix}
      0 & 1 \\
      1 & 0 
    \end{pmatrix}\,, 
    &
    \beta & =
    \begin{pmatrix}
      1 & 0 \\
      0 & -1 
    \end{pmatrix}\,.
  \end{aligned}
\end{equation}

In order to apply the Fourier split operator method to the Dirac
equation (\ref{eq:DE}), we split the Hamiltonian into an interaction
part and a free particle part
\begin{subequations}
  \begin{align}
    \label{eq:DE_V}
    {\I\hbar}\op{A}_1 
    & = c\sum_{i=1}^d
    \alpha_i\left( - q A_i(\vec x, t) \right) + q \phi(\vec x, t) \,,\\
    \label{eq:DE_K}
    {\I\hbar}\op{A}_2 
    & = c\sum_{i=1}^d
    \alpha_i\left(
      -\I\hbar\frac{\partial}{\partial r_i}
    \right) + mc^2\beta
  \end{align}
\end{subequations}
and calculate $\op {U}_{\op{A}_1}\left(t + \tau, t, \delta\right)$ and
$\op {\tilde U}_{\op{A}_2}\left(t + \tau, t, \delta\right)$ in the
following way.

The operator $\op {U}_{\op{A}_1}\left(t + \tau, t, \delta\right)$ may be
determined by splitting ${\op{A}_1}$ further into
\begin{equation}
  \op{A}_1 = \op{A}_{1,1} + \op{A}_{1,2}
\end{equation}
with
\begin{subequations}
  \begin{align}
    {\I\hbar}\op{A}_{1,1} & = q \phi(\vec x, t)\,, \\
    {\I\hbar}\op{A}_{1,2} & = 
    c\sum_{i=1}^d \alpha_i\left( - q A_i(\vec x, t) \right) \,.
  \end{align}
\end{subequations}
Because $\op{A}_{1,1}$ and $\op{A}_{1,2}$ commute we may factorize the
operator $\op
{U}_{\op{A}_1}\left(t + \tau, t, \delta\right)$ into
\begin{equation}
  \op {U}_{\op{A}_1}\left(t + \tau, t, \delta\right) =
  \op {U}_{\op{A}_{1,1}}\left(t + \tau, t, \delta\right)
  \op {U}_{\op{A}_{1,2}}\left(t + \tau, t, \delta\right)
\end{equation}
with the diagonal operator
\begin{equation}
  \op U_{\op{A}_{1,1}}\left(t+\tau, t, \delta\right)\Psi(\vec x, t) =
  \exp\left(
    -\delta\frac{\I}{\hbar}\int_t^{t+\tau} q \phi(\vec x, t') \,\D t'
  \right)\Psi(\vec x, t)\,.
\end{equation}
The operator $\op U_{\op{A}_{1,2}}\left(t+\tau, t, \delta\right)$,
however, involves matrix exponentials, which may be calculated by taking
into account the Dirac algebra (\ref{eq:Dirac_algebra}).  Exponentials
of $\alpha_i$ are obtained by summing the exponential-function's Taylor
sum explicitly. Introducing some auxiliary complex numbers $a_i$ we find
\begin{multline}
  \label{eq:exp_alpha}
  \exp\left(\I \sum_{i=1}^da_i\alpha_i\right) 
  = \sum_{k=0}^\infty \frac{1}{k!} \left(\I \sum_{i=1}^da_i\alpha_i\right)^k \\
  = \sum_{k=0}^\infty \frac{(-1)^k}{(2k)!} \left(\sum_{i=1}^da_i\alpha_i\right)^{2k} 
  +  \I \sum_{i=1}^da_i\alpha_i \sum_{k=0}^\infty \frac{(-1)^k}{(2k+1)!} \left(\sum_{i=1}^da_i\alpha_i\right)^{2k} \\ 
  = \sum_{k=0}^\infty \frac{(-1)^k}{(2k)!} \left(\sqrt{\sum_{i=1}^da_i^2}\right)^{2k} 
  +  \I \sum_{i=1}^da_i\alpha_i \sum_{k=0}^\infty \frac{(-1)^k}{(2k+1)!} \left(\sqrt{\sum_{i=1}^da_i^2}\right)^{2k} \\ 
  = \cos\left(\sqrt{\sum_{i=1}^da_i^2}\right) 
  +  \I \sum_{i=1}^da_i\alpha_i \left(\sum_{i=1}^da_i^2\right)^{-1/2}\sin\left(\sqrt{\sum_{i=1}^da_i^2}\right)\,,
\end{multline}
where we have used
\begin{multline}
  \left(\sum_{i=1}^da_i\alpha_i\right)^{2k} =
  \left(\left(\sum_{i=1}^da_i\alpha_i\right)^2\right)^k \\
  = \left(\sum_{i=1}^da_i^2\alpha_i^2 + \sum_{i=1}^d\sum_{j=1}^{i-1}a_ia_j(\alpha_i\alpha_j+\alpha_j\alpha_i)\right)^k =
  \left(\sqrt{\sum_{i=1}^da_i^2}\right)^{2k}\,.
\end{multline}
For convenience, let us define
\begin{subequations}
  \begin{align}
    \bar A_i(\vec x, t) & = \int_t^{t+\tau} A_i(\vec x, t')\,\D t' 
    \intertext{and}
    \bar A(\vec x, t) & = \sqrt{\sum_{i=1}^d\bar A_i(\vec x, t)^2} \,,
  \end{align}
\end{subequations}
then we get with (\ref{eq:exp_alpha}) 
\begin{multline}
  \op U_{\op{A}_{1,2}}\left(t+\tau, t, \delta\right)\Psi(\vec x, t) = \\
  \left(
    \cos\left(-\frac{\delta c }{\hbar}\bar A(\vec x, t)\right) +
    \I\sum_{i=1}^d\frac{\bar A_i(\vec x, t)}{\bar A(\vec x, t)}\alpha_i
    \sin\left(-\frac{\delta c }{\hbar}\bar A(\vec x, t)\right)
  \right)\Psi(\vec x, t) \,.
\end{multline}

The operator $\op U_{\op{A}_2}\left(t_2, t_1, \delta\right)$ equals the
time evolution operator of the free particle Dirac Hamiltonian. In
Fourier space it has the form
\begin{equation}
  \label{eq:U_A_2_DE}
  \op{\tilde U}_{\op{A}_2}\left(t+\tau, t, \delta\right) =
  \exp\left(
    -\delta\tau\frac\I\hbar
    \left(c\sum_{i=1}^d\alpha_i p_i + m c^2 \beta\right)
  \right)\,,
\end{equation}
where $p_i$ denotes the $i$th component of the momentum vector $\vec p$.
In order to calculate the operator exponential in (\ref{eq:U_A_2_DE}) we
have to diagonalize the operator
\begin{equation}
  {\I\hbar}\op{\tilde A}_2 = c\sum_{i=1}^d\alpha_i p_i + m c^2 \beta
\end{equation}
by introducing the scalars
\begin{equation}
  d_\pm(\vec p) =
  \left(\frac{1}{2} \pm \frac{1}{2\sqrt{1 + \vec p^2/(mc)^2}}\right)^{1/2} \,,
\end{equation}
the unitary matrix
\begin{equation}
  \label{eq:u_p}
  \op u(\vec p) = d_+(\vec p) + 
  d_-(\vec p)\sum_{i=1}^d \frac{p_i}{|\vec p|}\beta\cdot\alpha_i \,.
\end{equation}
and its Hermitian adjoint $u^\dagger(\vec p)$.  The matrix $\op u(\vec
p) {\I\hbar}\op{\tilde A}_2 u^\dagger(\vec p)$ is diagonal
\cite{Thaller:2000:AdvancedVisualQM} and it reads for two- or
three-dimensional systems
\begin{equation}
  \op u(\vec p)  {\I\hbar}\op{\tilde A}_2 \op u^\dagger(\vec p) = 
  \begin{pmatrix}
    E(\vec p) & 0 & 0 & 0 \\ 
    0 & E(\vec p) & 0 & 0 \\ 
    0 & 0 & -E(\vec p) & 0 \\ 
    0 & 0 & 0 & -E(\vec p)
  \end{pmatrix}\,,
\end{equation}
with
\begin{equation}
  E(\vec p) = \sqrt{m^2c^4 + \vec p^2c^2}\,.
\end{equation}
Thus, the Fourier space operator (\ref{eq:U_A_2_DE}) simplifies to
\begin{multline}
  \op{\tilde U}_{\op{A}_2}\left(t+\tau, t, \delta\right)\tilde\Psi(\vec p, t) = \\
  \op u(\vec p)^\dagger 
  \begin{pmatrix}
    \e^{-\I E(\vec p)\tau/\hbar}\hspace{-0.8ex} & \hspace{0.8ex}0 & \hspace{0.8ex}0 & \hspace{0.8ex}0 \\
    \hspace{0.8ex}0 & \e^{-\I E(\vec p)\tau/\hbar}\hspace{-0.8ex} & \hspace{0.8ex}0 & \hspace{0.8ex}0 \\
    \hspace{0.8ex}0 & \hspace{0.8ex}0 & \e^{\I E(\vec p)\tau/\hbar}\hspace{-0.8ex} & \hspace{0.8ex}0 \\
    \hspace{0.8ex}0 & \hspace{0.8ex}0 & \hspace{0.8ex}0 & \e^{\I E(\vec p)\tau/\hbar}
  \end{pmatrix}
  \op u(\vec p)\tilde\Psi(\vec p, t)\,.
\end{multline}

For the one-dimensional Dirac equation where the vector potential
reduces to a scalar $A_1(x, t)$ and the wave function $\Psi(x, t)$ has
only two components the operators $\op {U}_{\op{A}_{1,2}}\left(t + \tau,
  t, \delta\right)$ and $\op {U}_{\op{A}_2}\left(t + \tau, t,
  \delta\right)$ have to form
\begin{multline}
  \op U_{\op{A}_{1,2}}\left(t+\tau, t, \delta\right)\Psi(x, t) = \\
  \left(
    \cos\left(-\frac{\delta c }{\hbar}\bar A_1(x, t)\right) +
    \I\alpha_1
    \sin\left(-\frac{\delta c }{\hbar}\bar A_1(x, t)\right)
  \right)\Psi(x, t) 
\end{multline}
and
\begin{multline}
  \op{\tilde U}_{\op{A}_2}\left(t+\tau, t, \delta\right)\tilde\Psi(p, t) = \\
  \op u(p)^\dagger 
  \begin{pmatrix}
    \e^{-\I E(p)\tau/\hbar}\hspace{-0.8ex} & \hspace{0.8ex}0 \\
    \hspace{0.8ex}0 & \e^{\I E(p)\tau/\hbar}
  \end{pmatrix}
  \op u(p)\tilde\Psi(p, t)\,.
\end{multline}

\section{GPU implementations of the Fourier split operator method}
\label{sec:implementation}

\subsection{GPU computing}

As we have outlined in Section~\ref{sec:Fourier_split}, the Fourier
split operator method consists of three elementary steps, the
application of the two operators $\op U_{\op{A}_{1}}$ and $\op{\tilde
  U}_{\op{A}_2}$ plus the calculation of the fast Fourier transform and
its inverse. All these three steps may be parallelized.  The application
of $\op U_{\op{A}_{1}}$ or $\op{\tilde U}_{\op{A}_2}$ is embarrassingly
parallel---each grid point can be updated independently from
others---and also the fast Fourier transform may be parallelized
efficiently \cite{Govindaraju:etal:2008:FFT_GPU}.  

Scalable parallel performance, however, is difficult to attain on
traditional CPU systems due to the imbalance of high CPU speed and
relatively slow memory. The Fourier split operator method in particular
is inherently memory bandwidth bounded because the application of $\op
U_{\op{A}_{1}}$ or $\op{\tilde U}_{\op{A}_2}$ requires only $O(N)$
operations and the calculation of the Fourier transform takes $O(N\log
N)$, where $N$ denotes the number of grid points. Its low computational
complexity is a very beneficial feature of the Fourier split operator
method but it also means that there is no or little potential data reuse
making the memory bandwidth a limiting factor.

GPU computing
\cite{Kindratenko:etal:2010:_HPC_Accelerators,Owens:etal:2008:GPU_Computing}
may help to overcome these limitations.  Due to its massively parallel
architecture, GPUs reach a peak performance in floating point number
operations that is more than one order of magnitude higher than the peak
performance of current CPUs. GPUs also provide higher maximal memory
bandwidth. Intel's Core\texttrademark\,7 CPUs, for example, reach up to
25.6\,GB/s memory bandwidth while NVIDIA's Tesla\texttrademark\ M2050
and M2070 computing processors---which are based on GPU
technology---offer up to 148\,GB/s memory bandwidth, see also
Table~\ref{tab:GTX480_M2050}. Modern GPUs attain their impressive
computational performance figures without significantly exceeding the
power consumption of high-end CPUs. In fact, GPUs and other accelerators
provide the best floating point performance per watt of current high
performance computing architectures.

\subsection{CUDA parallel architecture and programming model}

GPU computing is enabled by programming models that provide a set of
abstractions that enable to express data parallelism and task
parallelism. These programming models are typically implemented by
equipping a sequential general purpose programming language, as for
example C or Fortran, with extensions for parallel programming and
providing an application programming interface. OpenCL \cite{OpenCL11},
Microsoft DirectCompute and CUDA by NVIDIA are the most popular
programming models for GPU computing. For our implementation of the
Fourier split operator method we chose the CUDA programming model
(version~3.2) which works only for graphics hardware by NVIDIA but
provides more flexibility than OpenCL or DirectCompute.  OpenCL is the
platform independent but has no support for C++ and DirectCompute works
only on Microsoft Windows systems.

The term CUDA also refers to some GPU architectures by the hardware
manufacturer NVIDIA. The latest CUDA GPU architecture is called
Fermi. It features up to 512 computing cores which are organized in 16
streaming multiprocessors of 32 cores each. Each of the 32~cores of a
streaming multiprocessor executes the same instruction on different data
sets at the same time.  Therefore, GPUs belong to the class \emph{Single
  Instruction, Multiple Data streams} (SIMD) in Flynn's taxonomy
\cite{Flynn:1972:taxonomy}.

\begin{table}
  \centering%
  \caption{Comparison of some technical key features of the high-end
    consumer graphics card NVIDIA GeForce GTX~480 and the computing processor
    module Tesla M2050 (source: NVIDIA).}
  \label{tab:GTX480_M2050}
  \begin{tabular}{@{}lrr@{}}
    \hline
    \hline
    & GeForce GTX 480 & Tesla M2050 \\
    \hline
    Processor cores & 480 & 448 \\
    Processor core clock & 1.40\,GHz & 1.15\,GHz \\
    Memory & 1.5\,GB & 3\,GB \\
    Memory clock & 1.848\,GHz & 1.546\,GHz \\
    Memory bandwidth & 177\,GB/s & 148\,GB/s \\
    Power consumption & $\null\approx 250$\,W & $\null\le 225$\,W \\
    \hline
    \hline
  \end{tabular}
\end{table}

GPU accelerator cards come in two different flavors, traditional
graphics cards, as the NVIDIA GeForce GTX~480, and dedicated computing
processors, as the NVIDIA Tesla M2050. Table~\ref{tab:GTX480_M2050}
compares some technical key features of these two cards. Both are based
on the so-called Fermi architecture but differ, for example, in memory
and clock rates.  According to this table, the dedicated computing
processor M2050 seems to be inferior to the GTX~480. However there are
two important features that qualify the M2050 for high performance
computing applications. It has larger memory with error correction and
full double precision performance. In consumer Fermi GPUs as the
GTX~480, the number of double precision floating point operations that
may be carried out per clock cycle is reduced by a factor of four as
compared to dedicated GPU computing processors based on the Fermi
architecture.

\begin{figure*}
  \hfill
  \includegraphics[scale=0.475]{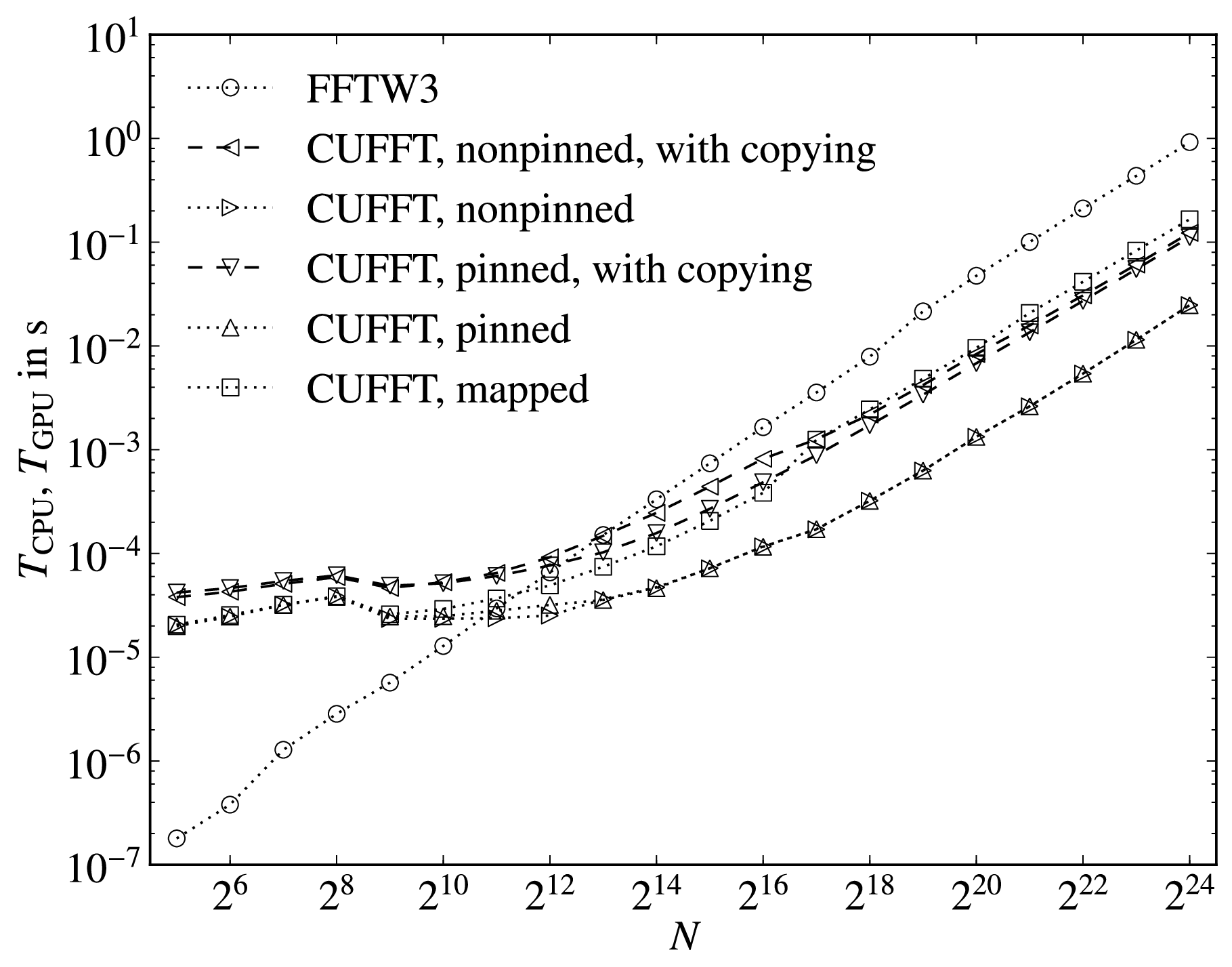}%
  \hfill\hfill%
  \includegraphics[scale=0.475]{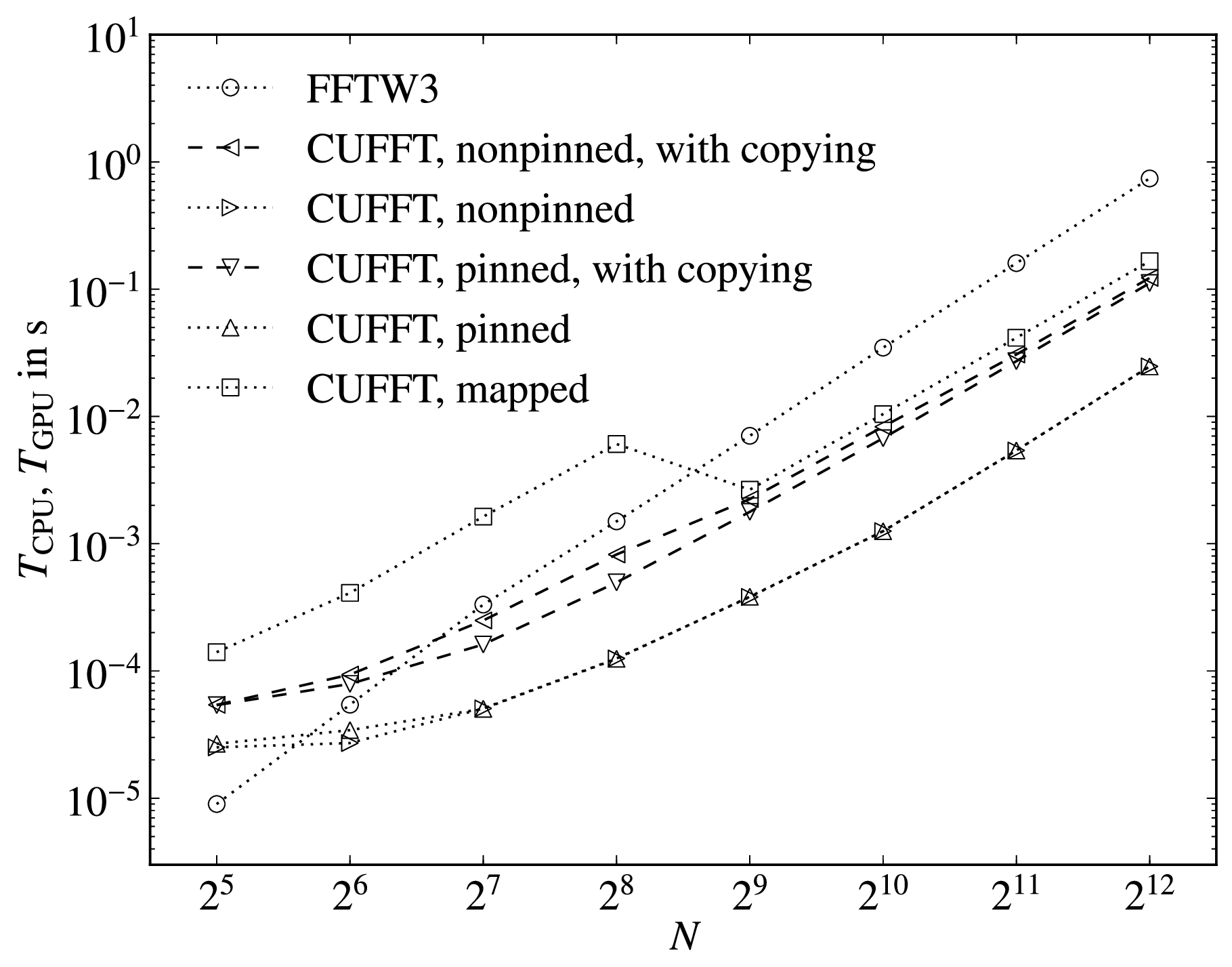}%
  \hfill\mbox{}%
  \caption{Wall clock time to compute a one-dimensional (left panel) and
    two-dimensional (right panel) fast Fourier transform of
    complex-valued double precision arrays of size $N$ and $N\times N$,
    respectively. Computations are done in place, that is, the input data
    is overwritten by the output. Times $T_\mathrm{CPU}$ and
    $T_\mathrm{GPU}$ denote time for computation utilizing a single CPU
    core (Intel Core i7 CPU at 2.93\,GHz) and the FFTW3 library and a
    GeForce~GTX~480 GPU and the CUFFT library, respectively. In the GPU
    case, measurements were carried out including and excluding data
    transfer between host memory (nonpinned or pinned) and GPU
    memory. Mapped memory resides on the host eliminating the need for
    data transfer.}
  \label{fig:FFT_perf}
\end{figure*}

A GPU has its own memory. This means, before one can run a GPU
computation the input data has to be transferred from the computer's
main memory (also called host memory) to the GPU device memory. GPU
computations are carried out in device memory and final results will be
transferred back into host memory. Memory transfer is a relatively slow
operation. Therefore, one should reduce the number of memory transfers
to reach high performance. The CUDA architecture provides three
different kinds of host memory, nonpinned, pinned, and mapped
memory. Memory transfers with pinned memory are faster than memory
transfers with nonpinned memory. However, allocating pinned memory is
slow. Mapped memory allows a GPU to work directly on data in host
memory. Thus, with mapped memory there is no need to copy data but a GPU
will access data in mapped memory not as fast as in device memory. We
will study the impact of memory transfer in
section~\ref{sec:performance_fft}.

\subsection{Implementation}

We have developed highly tuned codes for the Fourier split operator
method that allow us to propagate Schr{\"o}dinger wave functions and
Dirac wave functions in one and two dimensions. For each equation and
each dimension we implemented a conventional non-parallel CPU code as
well as a CUDA code that performs all computations on a GPU. To carry
out the fast Fourier transform CPU codes utilize the FFTW3 library
\cite{Frigo:Johnson:2005:FFTW3}. The GPU codes employ the CUFFT library
for fast Fourier transforms and comprise light-wight kernels that
accomplish the action of the operators $\op U_{{\op A}_1}$ and
$\op{\tilde U}_{{\op A}_2}$ for each grid point in parallel.

\section{Performance results}
\label{sec:performance}

\subsection{Fast Fourier transform}
\label{sec:performance_fft}

The overall performance of the Fourier split operator method is mainly
determined by the performance of the fast Fourier transform. Thus, we
compare in Figure~\ref{fig:FFT_perf} the wall clock time
$T_\mathrm{CPU}$ that is required to calculate the fast Fourier
transform on a CPU with the time $T_\mathrm{GPU}$ to perform the same
task on a GPU. For not too small problems the GPU outperforms the CPU
significantly. With our hardware setup (Intel Core~i7 CPU at 2.93\,GHz
with a GeForce~GTX~480 GPU) we got a speedup up to a factor of
about~30. For small problems, however, the overhead of starting and
coordinating parallel CUDA threads prevents to archive good fast Fourier
transform performance on GPUs. There is a critical problem size where
the GPU starts to outperform a CPU in calculating a fast Fourier
transform, which is for our hardware setup at data sets of about
$2^{12}$ complex numbers.

For GPU computations, one may use nonpinned, pinned, or mapped memory on
the host. If one includes the time to copy data from host memory to
device memory and back again after the calculation of the Fourier
transform in the measurements then the time depends on the kind of
memory. Pinned memory is faster than nonpinned memory. Mapped memory
eliminates the need to copy data between host and device, however,
performing the fast Fourier transform on device memory plus copying is
usually faster than calculating it in mapped host memory without data
transfer. Only for one-dimensional Fourier transforms of intermediate
size carrying out the fast Fourier transform in mapped host memory
requires less time than performing the same task on device memory plus
data transfer or using a (nonparallel) CPU implementation.

We took also fast Fourier transform performance measurements for a Tesla
M2050 computing processor. Despite the fact that the M2050 can perform
four times more double precision operations per clock cycle than the
consumer GPU GeForce~GTX~480, the Fourier transform performance of the
Tesla M2050 is slightly lower than for the GeForce~GTX~480. For a
two-dimensional Fourier transform of a $4096\times 4096$ array, for
example, the Tesla M2050 reached only about 95\,\% of the
GeForce~GTX~480 GPU performance. For smaller arrays the performance
difference was even larger. This may be interpreted as that the fast
Fourier transform performance is bound by the GPU's memory bandwidth
rather than by its floating point performance.

\begin{figure*}
  \hfill
  \includegraphics[scale=0.475]{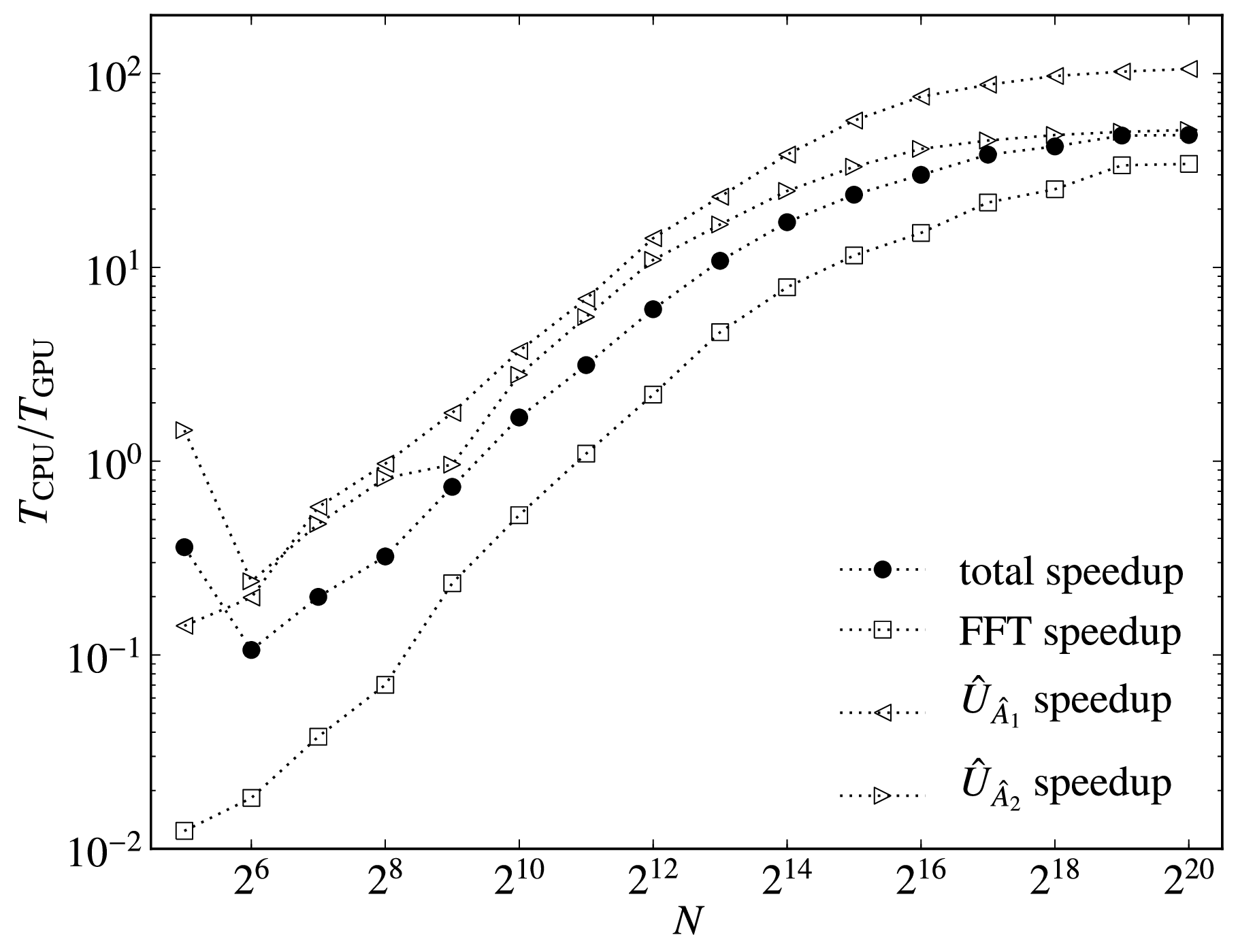}%
  \hfill\hfill%
  \includegraphics[scale=0.475]{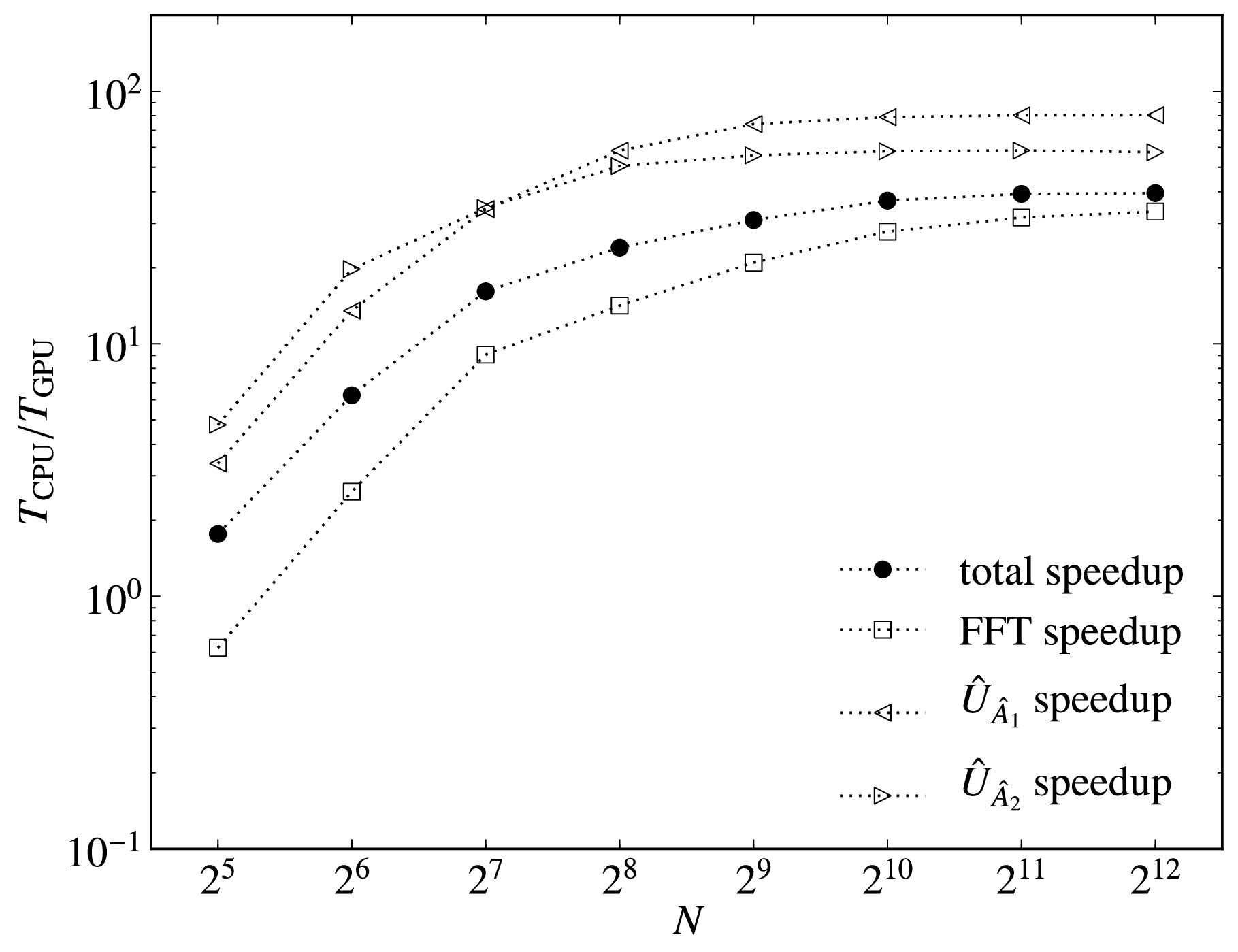}%
  \hfill\mbox{}%
  \caption{Speedup $T_\mathrm{CPU}/T_\mathrm{GPU}$ for propagating a
    wave function over 128 time steps as a function of the grid
    size. The left panel depicts results for the one-dimensional
    Schr{\"o}dinger equation of a grid of $N$ data points while the right
    panel shows results for the two-dimensional Dirac equation of a grid
    of $N\times N$ data points. Performance measurements carried out on
    the same hardware as in Figure~\ref{fig:FFT_perf} and in double
    precision.}
  \label{fig:speedup}
\end{figure*}

Our findings for the fast Fourier transform performance are consistent
with the results presented in the literature. Reference
\cite{Surkov:2010:Parallel_option_pricing} reports a performance gain up
to a factor of about 11 in calculating two-dimensional single precision
fast Fourier transforms on a NVIDIA GeForce~9800~GX2 GPU\footnote{Two
  hardware generations older than the GeForce~GTX~480 that we used.}
instead on a conventional CPU. In \cite{Govindaraju:etal:2008:FFT_GPU} a
8- to \mbox{40-fold} improvement was archived over highly tuned CPU
routines by using an algorithm that efficiently exploits GPU shared memory.

To sum up, the CUFFT GPU implementation of the fast Fourier transform is
capable to outperform traditional CPU implementations by more than one
order of magnitude. Data transfer between host memory and device memory
poses a non-negligible overhead and should be avoided if possible or
reduced by using pinned memory.

\subsection{Fourier split operator method}

In order to determine the speedup that may be attained by switching from
a CPU implementation to a GPU implementation of the Fourier split
operator method, we propagated one- and two-dimensional Gaussian wave
packets in a harmonic potential over 128 time steps under the effect of
the Schr{\"o}dinger equation (\ref{eq:SE}) and the Dirac equation
(\ref{eq:DE}), receptively. We measured
\begin{itemize*}
\item the overall time to propagate 128 time steps (In the case of GPU
  computations, this includes data transfer between host memory and
  device memory before the first step and after the last step.),
\item the time to perform the fast Fourier transform plus its inverse,
\item the time to apply the operator $\op U_{\op A_1}$ in position
  space, and
\item the time to apply the operator $\op{\tilde U}_{\op A_2}$ in
  Fourier space
\end{itemize*}
for various grids of size $N$ and $N\times N$. Figure~\ref{fig:speedup}
shows the speedup $T_\mathrm{CPU}/T_\mathrm{GPU}$ for these four
different tasks for the one-dimensional Schr{\"o}dinger equation and for the
two-dimensional Dirac equation as a function of the grid size. If one
considers the individual steps of the Fourier split operator method,
then one finds that the speedup for the application of $\op U_{\op A_1}$
and $\op{\tilde U}_{\op A_2}$ reaches between 50 to about 100 for large
systems (more than about $2^{18}$ grid points). For the fast Fourier
transform, however, we get a speedup of only about 30 limiting the
overall speedup to about 40 for large systems, which is still a
significant improvement over the CPU implementation. Results for the
two-dimensional Schr{\"o}dinger equation and the one-dimensional Dirac
equation are not shown in Figure~\ref{fig:speedup} because they are
qualitatively very similar to the two-dimensional Dirac equation and the
one-dimensional Schr{\"o}dinger equation, respectively.

\section{Applications}
\label{sec:applications}

In this section we will show some applications of our Fourier split
operator GPU codes for the solution of the time dependent 
Dirac equation. With conventional CPU codes
\cite{Mocken:Keitel:2008:FFT-split-operator}, these kinds of
applications would require more than an order of magnitude more
computing time or may not be performed in an admissible amount of
time. For our numerical simulation we adopt the atomic unit system
(a.\,u.) that is established on the Bohr radius, the electron's mass,
the reduced Planck constant and the absolute value of the electron's
charge as its base units.

\subsection{Evolution of a free Gaussian wave packet}
\label{subsec:free_Gaussian_wave_packet}

A $d$-dimensional wave packet may be formed by superimposing plane waves
solutions of the Dirac equation with defined momentum $\vec p$, viz.
\begin{equation}
  \Psi_{\mathrm{packed}}(\vec x) = 
  \frac{1}{(2\pi\hbar)^{d/2}}
  \int \rho(\vec p)
  \exp\left(
    \frac{\I \vec x\cdot\vec p}{\hbar}
  \right)\,\D^d p
\end{equation}
The quantity $\rho(\vec p)$ determines the momentum distribution. For a
two-dimensional Gaussian distribution with mean momentum $\bar{\vec p}$
and momentum widths $\sigma_f$ in forwards direction and $\sigma_s$ in
sidewards direction the function $\rho(\vec p)$ reads
\begin{equation}
  \rho_{\mathrm{Gaussian}}(\vec p) =
  u(\vec p)\frac{1}{(2\pi)^{1/2} |\Sigma|^{1/4}}
  \exp\left(
    -\frac{(\vec p-\bar{\vec p})^T\Sigma^{-1}(\vec p-\bar{\vec p})}{4}
  \right)\,,
\end{equation}
where $u(\vec p)$ denotes a column of the matrix $\tilde u(\vec p)$
(\ref{eq:u_p}) (selecting one the four momentum eigenstates with
momentum $\vec p$) and with the matrix
\begin{equation}
  \Sigma^{-1} = R
  \begin{pmatrix}
    1/\sigma_f^2 & 0 \\ 0 & 1/\sigma_s^2
  \end{pmatrix}
  R^{-1}
\end{equation}
and 
\begin{equation}
  R =
  \begin{pmatrix}
    \cos(\phi) & -\sin(\phi) \\ \sin(\phi) & \cos(\phi)
  \end{pmatrix}
  \qquad\text{and}\qquad
  \tan \phi = \frac{\bar p_y}{\bar p_x}\,.
\end{equation} 

\begin{figure}
  \centering
  \includegraphics[scale=0.475]{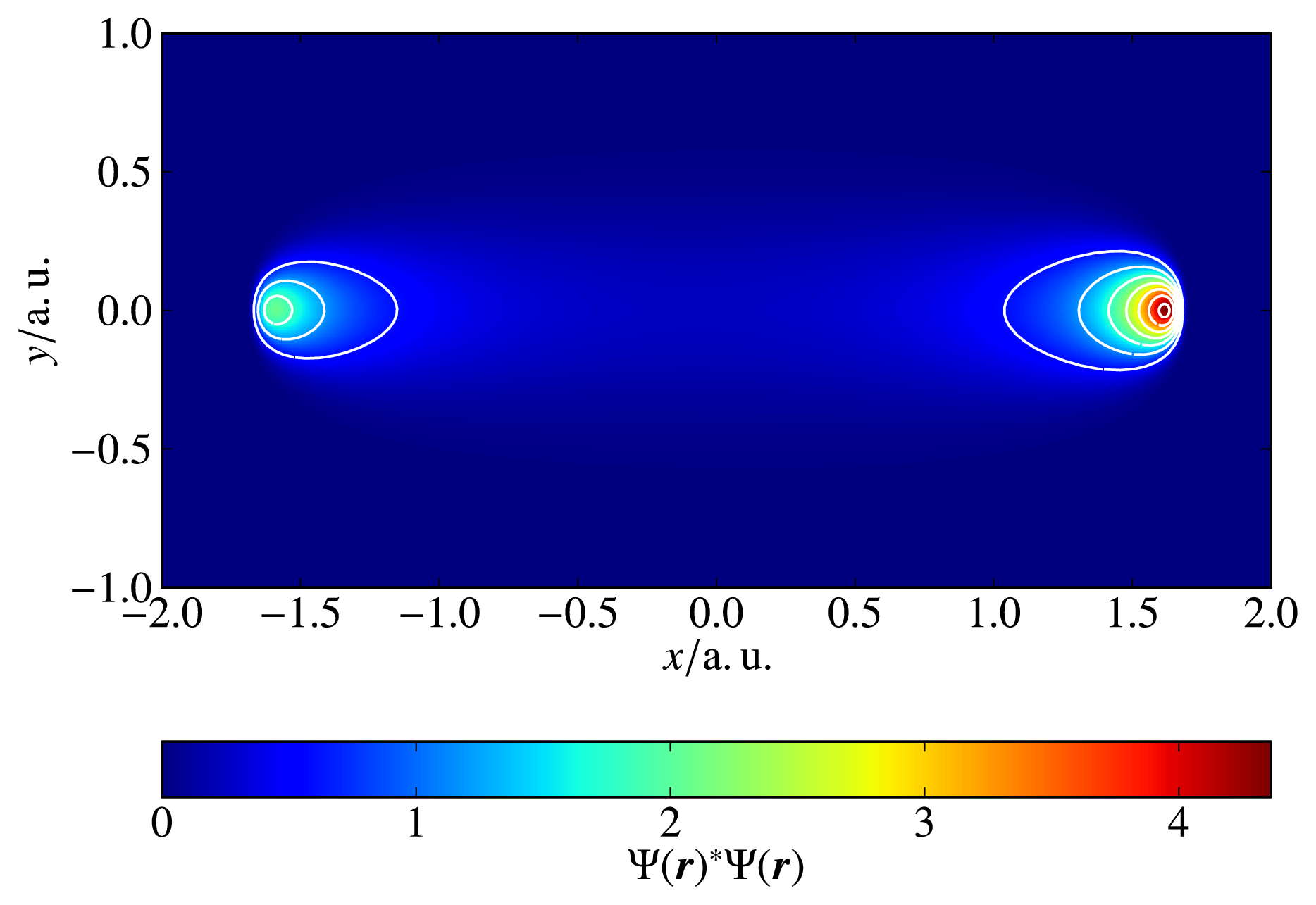}
  \caption{Probability density of a Dirac wave packed with narrow
    asymmetric momentum distribution at time $t=0.0125\,\mbox{a.\,u.}$,
    see text for specific parameters. Due to the finite speed of light,
    the wave packet splits into two shock fronts traveling into opposite
    directions.}
  \label{fig:free_particle_2d}
\end{figure}

The evolution of the probability density of a relativistic Dirac wave
packet differs significantly from the non-relativistic theory of the
Schr{\"o}dinger equation, where a Gaussian wave packet broadens but
remains its Gaussian shape for all times and the maximum of the
probability density travels with the same speed as the motion of the
center of mass does.  We consider the propagation of a relativistic
Gaussian Dirac wave packet in two dimensions with asymmetric momentum
widths $\sigma_f=200\,\mbox{a.\,u.}$ and $\sigma_s=20\,\mbox{a.\,u.}$, a
mean momentum $\bar{\vec p}=(40\,\mbox{a.\,u.},0\,\mbox{a.\,u.})$,
positive energy and spin up. This corresponds to a rather narrow initial
wave packet. The probability density of this two-dimensional
relativistic Dirac wave packet at time $t=0.0125\,\mbox{a.\,u.}$ is
illustrated in \figurename~\ref{fig:free_particle_2d}. The position of
the initial probability density's maximum corresponds to the initial
center of mass at the center of the coordinate system.  The wave
packet's center of mass moves in accordance with classical predictions
with velocity $|\bar{\vec p}|/(m\sqrt{1+\bar{\vec p}^2/(mc)^2})\approx
38\,\mbox{a.\,u.}$ along the $x$-axis.  Due to the finite speed of light
\cite{Thaller:2004:Advanced_Visual_QM}, however, shock fronts emerge
traveling approximately with the speed of light into the forward and
backward directions. The maxima of the probability density no longer
coincide with the center of mass. See
\cite{Ruf:2009:Klein-Gordon_Splitoperator,Su:etal:1998:Relativistic_suppression}
for an investigation of similar relativistic effects.

\subsection{Eigenstates}

Propagating a trail wave function $\Psi(\vec x, t)$ in a time
independent potential from $t=0$ to some later time $t=T$ allows us to
determine the potential's eigenenergies
\cite{Feit:etal:1982:spectral_method} by calculating the autocorrelation
function
\begin{equation}
  \chi(t) = \int \Psi(\vec x, 0)^*\Psi(\vec x, t)\,\D^dx\,.
\end{equation}
The Fourier transform of the autocorrelation function
\begin{equation}
  \tilde\chi(E) = 
  \int_0^T \chi(t)(1-\cos(2\pi t/T))\exp(\I tE/\hbar) \,\D t
\end{equation}
exhibits pronounced peaks at the eigenenergies provided that the initial
trial function is not orthogonal to some eigenfunction. Once we have
determined an eigenenergy $E$ from the spectrum $\tilde\chi(E)$, the
potential's eigenfunction $\Psi_E(\vec x)$ (assuming there is no
degeneracy) with eigenenergy $E$ results from the integral
\begin{equation}
  \Psi_E(\vec x) \sim 
  \int_0^T \Psi(\vec x, t)(1-\cos(2\pi t/T))\exp(\I Et/\hbar)\,\D t \,.
\end{equation}

\begin{figure}
  \centering
  \includegraphics[scale=0.475]{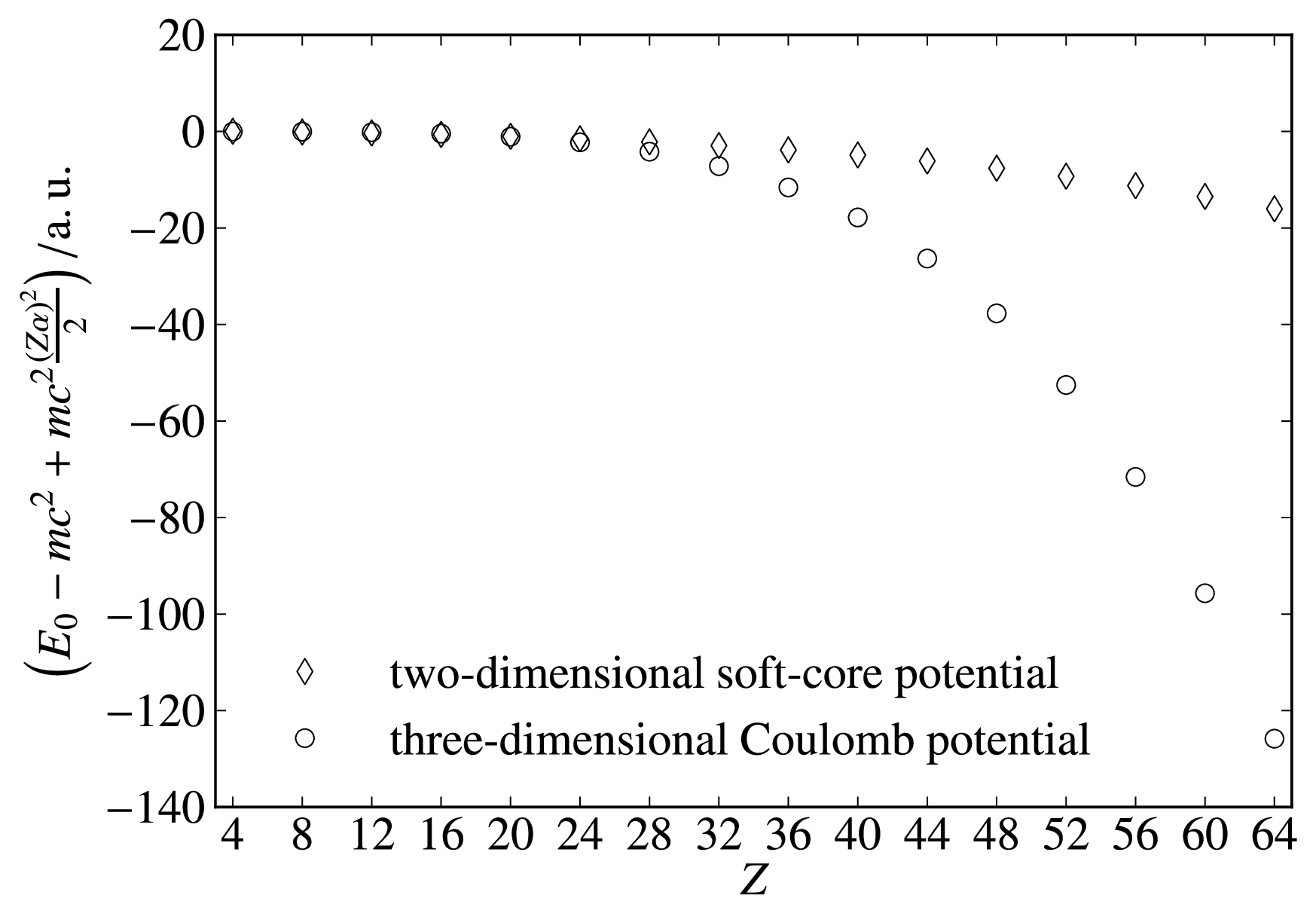}
  \caption{Relativistic corrections to the ground state energy $E_0$ for
    the two-dimensional soft-core potential (\ref{eq:soft_core}) with
    $\zeta=0.791$ and the three-dimensional soft-core potential for
    different values of the atomic number $Z$. In the non-relativistic
    Schr{\"o}dinger theory, both potentials share approximately the same
    ground state energy of $-mc^2(\alpha Z)^2/2$.}
  \label{fig:soft_core_spectrum_DE_2d}
\end{figure}

For one- or two-dimensional model systems it is common practice to mimic
the Coulomb potential
\begin{equation}
  \label{eq:Coulomb}
  V_\mathrm{C}(\vec x) = -e\phi(\vec x) = 
  -\frac{Ze^2}{4\pi\varepsilon_0|\vec x|}
\end{equation}
by a soft-core potential
\begin{equation}
  \label{eq:soft_core}
  V_\mathrm{sc}(\vec x) = -e\phi(\vec x) = 
  -\frac{Ze^2}{4\pi\varepsilon_0\sqrt{\vec x^2 + (\zeta/Z)^2}}\,.
\end{equation}
In (\ref{eq:Coulomb}) and (\ref{eq:soft_core}) $e$ denotes the
elementary charge, $Z$ the atomic number, $\varepsilon_0$ the vacuum
permittivity, and $\zeta$ the soft-core parameter.

If one replaces the three-dimensional Coulomb potential
(\ref{eq:Coulomb}) by a lower dimensional soft-core potential
(\ref{eq:soft_core}), it is often appropriate to choose the soft-core
parameter $\zeta$ such that both potential share the same groundstate
energy. In Schr{\"o}dinger theory, the Coulomb potential
(\ref{eq:Coulomb}) has a groundstate energy of $-mc^2(\alpha Z)^2/2$,
where we have introduced the electron mass $m$ and the fine structure
constant $\alpha=e^2/(4\pi\varepsilon_0\hbar c)$.  Numerically we find
that for $\zeta\approx1.4133$ the one-dimensional soft-core potential
has approximately the same groundstate energy as the three-dimensional
Coulomb potential with the same atomic number. In two dimensions, one
has to set $\zeta\approx 0.791$ in order to mach the groundstate
energies. These statements hold for all $Z$.

\begin{figure*}
  \centering
  \includegraphics[scale=0.475]{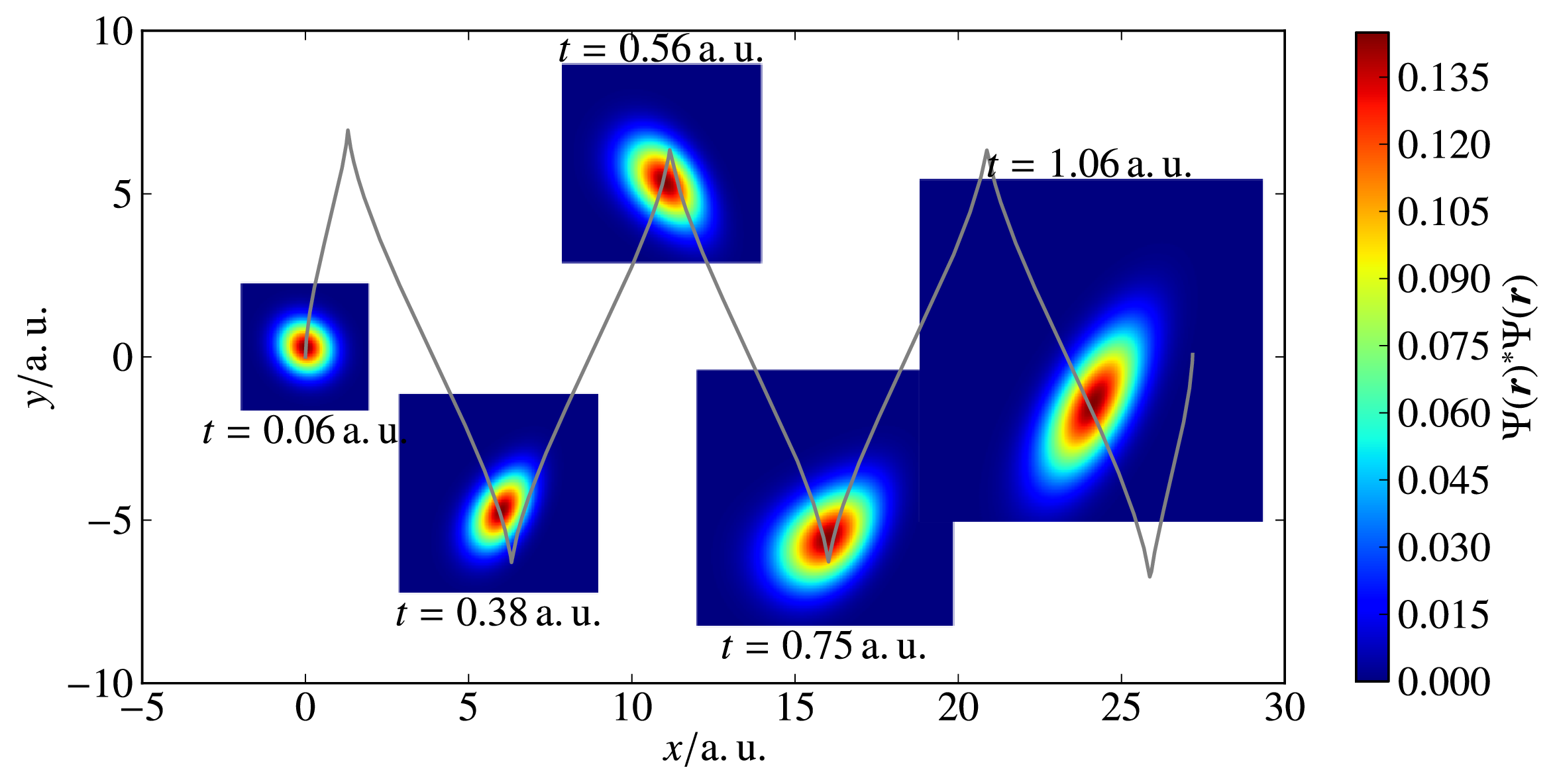}
  \caption{Free wave packed scatting at a strong laser pulse. The false
    color plots show the wave-packet's probability density at different
    points in time $t$. The solid gray line indicates the center of mass
    trajectory. The laser pulse travels from left to right. See text for
    detailed parameters. Note that the computational grid follows the
    center of mass motion.}
  \label{fig:free_particle_laser_DE_2d}
\end{figure*}

In Dirac's quantum theory, the groundstate energy of the
three-dimensional Coulomb potential (\ref{eq:Coulomb}) is
$mc^2\sqrt{1-(\alpha Z)^2}$ and the soft-core parameter that matches the
groundstate energies depends on $Z$, however, it is close to the values
for the Schr{\"o}dinger case. The Dirac groundstate energy of the Coulomb
potential equals the Schr{\"o}dinger groundstate energy plus the rest mass
energy $mc^2$ and relativistic corrections proportional in $Z^4$, viz.
\begin{equation}
  mc^2\sqrt{1-(\alpha Z)^2} =
  mc^2 - 
  mc^2 \frac{(\alpha Z)^2}{2} -
  mc^2 \frac{(\alpha Z)^4}{8} + \dots 
\end{equation}
We determined the Dirac groundstate energy $E_0$ of the two-dimensional
soft-core potential (\ref{eq:soft_core}) with $\zeta=0.791$ as a
function of the atomic number $Z$ with high numerical accuracy and found
that relativistic corrections to the two-dimensional
soft-core-potential's groundstate energy are weaker than for the
three-dimensional Coulomb potential as shown in
Fig.~\ref{fig:soft_core_spectrum_DE_2d}.

\subsection{Free wave packed scatting at a light pulse}

At high laser intensities, charged particles in electromagnetic waves
are not only affected by the electric field component but also by the
magnetic field component. At intensities larger than about
$55m\hbar\omega^2/(c\mu_0q^2)$, the Lorentz force becomes relevant and
at intensities above $0.1m^2c^2\omega^2/q^2$ also relativistic effects
have to be taken into account \cite{Reiss:2000:Dipole-approximation},
here $\omega$ denotes the laser's angular frequency and $\mu_0$ the
vacuum permeability.

We simulated the two-dimensional motion of a free wave packed in a laser
pulse with the electromagnetic fields
\begin{subequations}
  \label{eq:laserpulse}
  \begin{align}
    \vec E(\vec x, t) & = 
    \vec E_0 \sin(\vec k\cdot\vec x - \omega t)
    f_{j, l}(\vec k\cdot\vec x - \omega t)\,, \\
    \vec B(\vec x, t) & = 
    \frac{\vec E_0}{c} \sin(\vec k\cdot\vec x - \omega t) 
    f_{j, l}(\vec k\cdot\vec x - \omega t)
    \intertext{with an envelope function $f_{j, l}(\eta)$ of $j$ half
      cycles with a linear turn-on ramp and a linear turn-off ramp of
      $l$ half cycles and a constant plateau in between, viz.}
    f_{j, l}(\eta) & = \begin{cases}
      (\eta+j\pi)/(l\pi) & \text{if $-j\le\eta/\pi\le-j+l$,} \\
      1 & \text{if $-j+l\le\eta/\pi\le-l$,} \\
      -\eta/(l\pi) & \text{if $-l\le\eta/\pi\le0$,} \\
      0 & \text{else}
    \end{cases}
  \end{align}
\end{subequations}
and $|\vec k|=2\pi/\lambda=\omega/c$ and $\vec E_0\perp\vec
k$. Figure~\ref{fig:free_particle_laser_DE_2d} shows the wave-function's
probability density at different times for a free wave packed scatting
at a strong laser pulse traveling along the \mbox{$x$-direction} and
having an overall length of eight half cycles and turn-on/off ramps of
two half cycles. Its field strength is $|\vec E_0|=3000\,\mbox{a.\,u.}$
and its wave length $\lambda=40\,\mbox{a.\,u.}$ The electric field
component accelerates the wave packet back and forth along the
\mbox{$y$-direction}, while the Lorenz force causes a drift into the
\mbox{$x$-direction}.

\subsection{Ionization}

\begin{figure}[bt]
  \centering
  \includegraphics[scale=0.475]{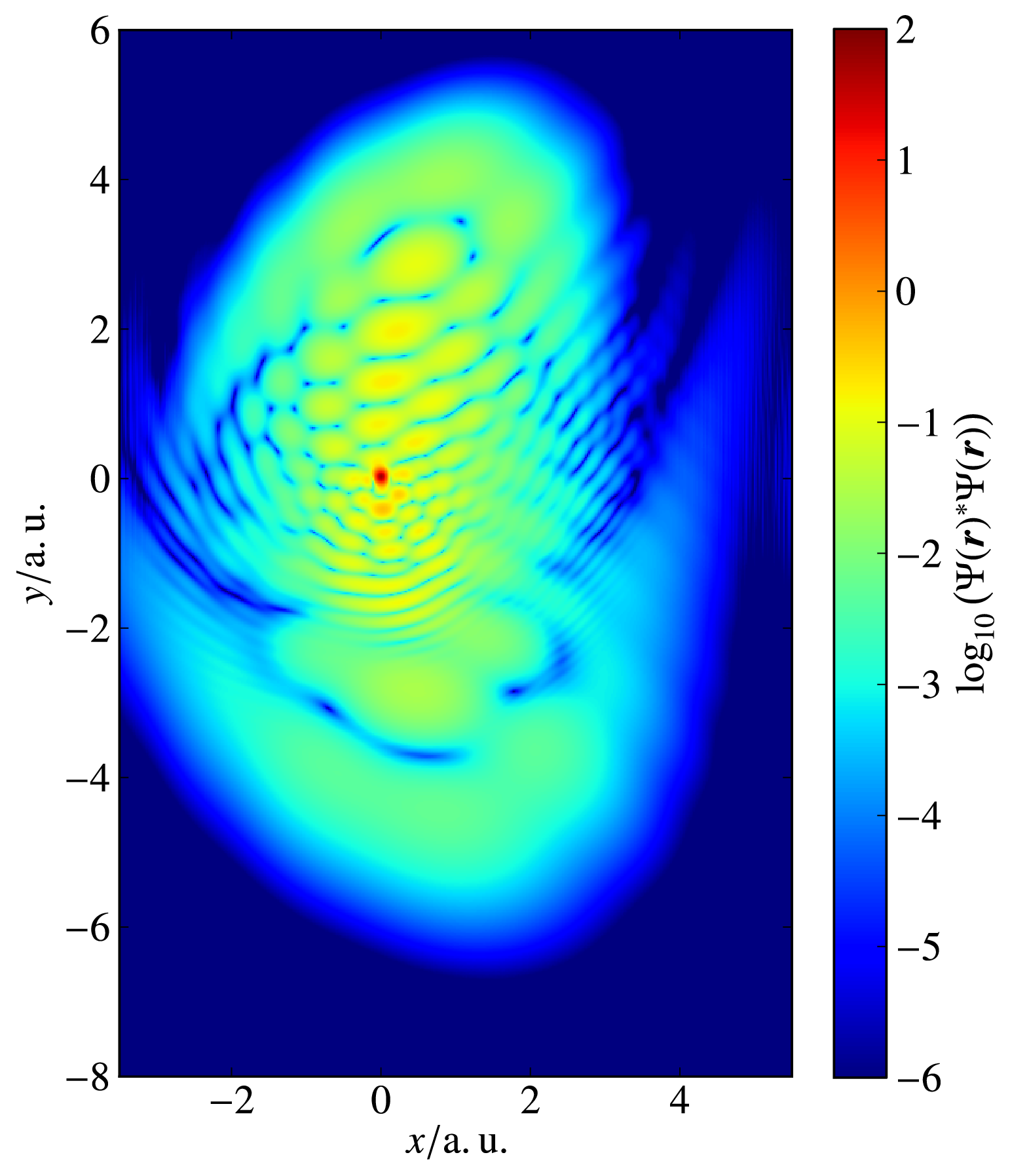}
  \caption{A wave-function's probability density after ionization from
    the ground state of a soft-core potential (\ref{eq:soft_core}) by a
    ultra-strong laser pulse, see text for details.}
  \label{fig:soft_core_DE_ionization}
\end{figure}

In our fourth example, we consider ionization in ultra-strong and
ultra-short laser pulses. The Dirac ground state wave packet of a
soft-core potential (\ref{eq:soft_core}) with $Z=32$ and $\zeta=0.791$
is excited by an external laser pulse (\ref{eq:laserpulse}) with
wavelength $\lambda=10\,\mbox{a.\,u.}$ and peak electric field strength
$|\vec E_0|=3072\,\mbox{a.\,u.}$. We consider a short laser pulse of
four half-cycles including a tun-on and a turn-off
half-cycle. Figure~\ref{fig:soft_core_DE_ionization} shows the electron
density after the laser pulse has passed the atomic core. In the setup,
we have chosen here, the laser travels from left to right and the
electric field points into the $y$-direction, accelerating the wave
packet into the positive $y$-direction in the first and third
half-cycles and into the negative $y$-direction during the second and
fourth half-cycles. The Lorenz force causes an acceleration into the
laser's propagation direction. The strong interference patters in
Fig.~\ref{fig:soft_core_DE_ionization} form by the interaction of
wave-function's different components traveling into different
directions.

Note the huge wave function spreading during the ionization. While the
ground state wave function has a width of about $1/32\,\mbox{a.\,u.}$,
the wave function after ionization spreads over several atomic units as
shown in Fig.~\ref{fig:soft_core_DE_ionization}. Wave function spreading
poses a major challenge in the numerical simulation of light matter
interaction at high intensities limiting the simulations to very short
time scales where the size of the wave function remains of manageable
size. The simulation of the ionization process of
Fig.~\ref{fig:soft_core_DE_ionization} to only about ten minutes using
our GPU Fourier split operator implementation. Thus, the simulation of
even longer interactions with laser pulses of longer wave lengths has
become feasible by our GPU implementation.

\section{Conclusions and outlook}
\label{sec:conclusions}

In this contribution we evaluated GPUs as a massively parallel computing
architecture for the solution of time dependent partial differential
equations by means of the Fourier split operator method. The
computational complexity of the Fourier split operator method is
dominated by the computational complexity of the fast Fourier
transform. We demonstrated that GPUs reach much better performance in
computing the fast Fourier transform than current CPUs. Depending on the
problem size, the performance gain may exceed one order of magnitude as
compared to sequential CPU implementations. Thus, the Fourier split
operator method may be implemented very efficiently on GPU architectures
as we demonstrated for the time dependent Schr{\"o}dinger equation and
the time dependent Dirac equation.  The combination of a highly parallel
architecture with a high-throughput memory makes graphics processing
units a very attractive architecture for implementing the Fourier split
operator method. Best performance is attained if all steps of the
Fourier split operator method are carried out by the GPU avoiding data
transfer between host memory and GPU memory.

The fast Fourier transform is a core building block for the solution of
partial differential equations as well as of many other problems from
computational physics, signal processing, tomography, computational
finance and other fields.  Thus, we expect that also these problems may
be solved much more efficiently on GPU architectures than on
conventional CPUs. Taking into account that GPU computing is a rather
young field it is supposed that there is still plenty potential for GPU
technology and codes to mature further and to find new applications.

\section*{Acknowledgments}

We would like to thank Martin Wolf and Sven Ahrens for stimulating
discussions.

\bibliographystyle{elsarticle-num}
\bibliography{gpu_split_operator}

\begin{thebibliography}{10}
\expandafter\ifx\csname url\endcsname\relax
  \def\url#1{\texttt{#1}}\fi
\expandafter\ifx\csname urlprefix\endcsname\relax\def\urlprefix{URL }\fi
\expandafter\ifx\csname href\endcsname\relax
  \def\href#1#2{#2} \def\path#1{#1}\fi

\bibitem{Kirk:Hwu:2010:Programming_MPP}
D.~Kirk, W.~Hwu, Programming Massively Parallel Processors: A Hands-on
  Approach, Morgan Kaufmann, 2010.

\bibitem{Kindratenko:etal:2010:_HPC_Accelerators}
V.~Kindratenko, R.~Wilhelmson, R.~Brunner, T.~J. Mart\"{i}nez, W.~Hwu,
  High-performance computing with accelerators, Computing in Science \&
  Engineering 12~(4) (2010) 12--16.
\newblock \href {http://dx.doi.org/10.1109/MCSE.2010.88}
  {\path{doi:10.1109/MCSE.2010.88}}.

\bibitem{Owens:etal:2008:GPU_Computing}
J.~Owens, M.~Houston, D.~Luebke, S.~Green, J.~Stone, J.~Phillips, {GPU}
  computing, Proceedings of the IEEE 96~(5) (2008) 879--899.
\newblock \href {http://dx.doi.org/10.1109/JPROC.2008.917757}
  {\path{doi:10.1109/JPROC.2008.917757}}.

\bibitem{Sutter:2005:free_lunch}
H.~Sutter, A fundamental turn toward concurrency in software, Dr. Dobb's
  Journal 30~(3) (2005) 202--210.

\bibitem{Risken:1989:Focker-Planck}
H.~Risken, The {Fokker}-{Planck} Equation, 2nd Edition, Vol.~18 of Springer
  Series in Synergetics, Springer, 1989.

\bibitem{Harshawardhan:etal:2000:Maxwell}
W.~Harshawardhan, Q.~Su, R.~Grobe, Numerical solution of the time-dependent
  {Maxwell's} equations for random dielectric media, Physical Review E 62~(6)
  (2000) 8705--8712.
\newblock \href {http://dx.doi.org/10.1103/PhysRevE.62.8705}
  {\path{doi:10.1103/PhysRevE.62.8705}}.

\bibitem{Thaller:1992:Dirac}
B.~Thaller, The {Dirac} Equation, Texts and Monographs in Physics, Springer,
  1992.

\bibitem{Feshbach:Villars:1954:Relativistic_Wave_Mechanics}
H.~Feshbach, F.~Villars, Elementary relativistic wave mechanics of spin 0 and
  spin 1/2 particles, Reviews of Modern Physics 30~(1) (1958) 24--45.
\newblock \href {http://dx.doi.org/10.1103/RevModPhys.30.24}
  {\path{doi:10.1103/RevModPhys.30.24}}.

\bibitem{Pechukas:Light:1966:Time-Displacement}
P.~Pechukas, J.~C. Light, On the exponential form of time-displacement
  operators in quantum mechanics, The Journal of Chemical Physics 44~(10)
  (1966) 3897--3912.
\newblock \href {http://dx.doi.org/10.1063/1.1726550}
  {\path{doi:10.1063/1.1726550}}.

\bibitem{Fattorini:1985:Cauchy}
H.~O. Fattorini, The {Cauchy} Problem, Vol.~18 of Encyclopedia of Mathematics
  and its Applications, Cambridge University Press, 1985.

\bibitem{Fleck:etal:1976:laser_propagation}
J.~A. Fleck, J.~R. Morris, M.~D. Feit, Time-dependent propagation of high
  energy laser beams through the atmosphere, Applied Physics 10~(2) (1976)
  129--160.

\bibitem{Feit:etal:1982:spectral_method}
M.~D. Feit, J.~A. {Fleck, Jr.}, A.~Steiger, Solution of the {Schr\"odinger}
  equation by a spectral method, Journal of Computational Physics 47~(3) (1982)
  412--433.
\newblock \href {http://dx.doi.org/10.1016/0021-9991(82)90091-2}
  {\path{doi:10.1016/0021-9991(82)90091-2}}.

\bibitem{Bandrauk:Shen:1997:coupled_time-dependent_SE}
A.~D. Bandrauk, H.~Shen, Exponential split operator methods for solving coupled
  time-dependent {Schr\"odinger} equations, Journal of Chemical Physics 99~(2)
  (1993) 1185--1193.
\newblock \href {http://dx.doi.org/10.1063/1.465362}
  {\path{doi:10.1063/1.465362}}.

\bibitem{Braun:etal:1999:Numerical_approach}
J.~W. Braun, Q.~Su, R.~Grobe, Numerical approach to solve the time-dependent
  {Dirac} equation, Physical Review A 59~(1) (1999) 604--612.
\newblock \href {http://dx.doi.org/10.1103/PhysRevA.59.604}
  {\path{doi:10.1103/PhysRevA.59.604}}.

\bibitem{Mocken:Keitel:2004:Quantum_dynamics}
G.~R. Mocken, C.~H. Keitel, Quantum dynamics of relativistic electrons, Journal
  of Computational Physics 199~(2) (2004) 558--588.
\newblock \href {http://dx.doi.org/10.1016/j.jcp.2004.02.020}
  {\path{doi:10.1016/j.jcp.2004.02.020}}.

\bibitem{Mocken:Keitel:2008:FFT-split-operator}
G.~R. Mocken, C.~H. Keitel, {FFT}-split-operator code for solving the {Dirac}
  equation in $2+1$ dimensions, Computer Physics Communications 178~(11) (2008)
  868--882.
\newblock \href {http://dx.doi.org/10.1016/j.cpc.2008.01.042}
  {\path{doi:10.1016/j.cpc.2008.01.042}}.

\bibitem{Hu:Keitel:2001:Dynamics}
S.~X. Hu, C.~H. Keitel, Dynamics of multiply charged ions in intense laser
  fields, Physical Review~A 63~(5) (2001) 053402.
\newblock \href {http://dx.doi.org/10.1103/PhysRevA.63.053402}
  {\path{doi:10.1103/PhysRevA.63.053402}}.

\bibitem{Javanainen:Ruostekoski:2006:Symbolic_calculation}
J.~Javanainen, J.~Ruostekoski, Symbolic calculation in development of
  algorithms: split-step methods for the {Gross}-{Pitaevskii} equation, Journal
  of Physics A 39~(12) (2006) L179--L184.
\newblock \href {http://dx.doi.org/10.1088/0305-4470/39/12/L02}
  {\path{doi:10.1088/0305-4470/39/12/L02}}.

\bibitem{Muslu:Erbay:2005:Higher-order_split-step}
G.~M. Muslu, H.~A. Erbay, Higher-order split-step {Fourier} schemes for the
  generalized nonlinear {Schr\"odinger} equation, Mathematics and Computers in
  Simulation 67~(6) (2005) 581--595.
\newblock \href {http://dx.doi.org/10.1016/j.matcom.2004.08.002}
  {\path{doi:10.1016/j.matcom.2004.08.002}}.

\bibitem{Strang:1968:Difference_Schemes}
G.~Strang, On the construction and comparison of difference schemes, SIAM
  Journal on Numerical Analysis 5~(3) (1968) 506--517.
\newblock \href {http://dx.doi.org/10.1137/0705041}
  {\path{doi:10.1137/0705041}}.

\bibitem{Thaller:2000:AdvancedVisualQM}
B.~Thaller, Advanced Visual Quantum Mechanics, Springer, 2004.

\bibitem{Govindaraju:etal:2008:FFT_GPU}
N.~K. Govindaraju, B.~Lloyd, Y.~Dotsenko, B.~Smith, J.~Manferdelli, High
  performance discrete {Fourier} transforms on graphics processors, in:
  Proceedings of the 2008 {ACM/IEEE} conference on Supercomputing, Piscataway,
  NJ, USA, 2008, pp. 1--12.

\bibitem{OpenCL11}
Khronos OpenCL Working Group, The {OpenCL} Specification Version 1.1 (2010).

\bibitem{Flynn:1972:taxonomy}
M.~J. Flynn, Some computer organizations and their effectiveness, IEEE
  Transactions on Computers C-21~(9) (1972) 948--960.
\newblock \href {http://dx.doi.org/10.1109/TC.1972.5009071}
  {\path{doi:10.1109/TC.1972.5009071}}.

\bibitem{Frigo:Johnson:2005:FFTW3}
M.~Frigo, S.~G. Johnson, The design and implementation of {FFTW3}, Proceedings
  of the IEEE 93~(2) (2005) 216--231.
\newblock \href {http://dx.doi.org/10.1109/JPROC.2004.840301}
  {\path{doi:10.1109/JPROC.2004.840301}}.

\bibitem{Surkov:2010:Parallel_option_pricing}
V.~Surkov, Parallel option pricing with {Fourier} space time-stepping method on
  graphics processing units, Parallel Computing 36~(7) (2010) 372--380.
\newblock \href {http://dx.doi.org/10.1016/j.parco.2010.02.006}
  {\path{doi:10.1016/j.parco.2010.02.006}}.

\bibitem{Note3}
Two hardware generations older than the GeForce~GTX~480 that we used.

\bibitem{Thaller:2004:Advanced_Visual_QM}
B.~Thaller, Advanced Visual Quantum Mechanics, Springer, 2004.

\bibitem{Ruf:2009:Klein-Gordon_Splitoperator}
M.~Ruf, H.~Bauke, C.~H. Keitel, A real space split operator method for the
  {Klein-Gordon} equation, Journal of Computational Physics 228~(24) (2009)
  9092--9106.
\newblock \href {http://dx.doi.org/10.1016/j.jcp.2009.09.012}
  {\path{doi:10.1016/j.jcp.2009.09.012}}.

\bibitem{Su:etal:1998:Relativistic_suppression}
Q.~Su, B.~Smetanko, R.~Grobe, Relativistic suppression of wave packet
  spreading, Optics Express 2~(7) (1998) 277--281.

\bibitem{Reiss:2000:Dipole-approximation}
H.~R. Reiss, Dipole-approximation magnetic fields in strong laser beams,
  Physical Review A 63 (2000) 013409.
\newblock \href {http://dx.doi.org/10.1103/PhysRevA.63.013409}
  {\path{doi:10.1103/PhysRevA.63.013409}}.

\end{thebibliography}

\end{document}